\documentclass[aps,prl,12pt,reprint]{revtex4-1}

\usepackage{graphicx}
\usepackage{amsmath}
\usepackage{amssymb}

\newcommand*{\Temp}{\mathcal{T}}

\begin{document}

\title{High precision nano scale temperature sensing using single defects in diamond} 

\author{P. Neumann}
\email{p.neumann@physik.uni-stuttgart.de}
\affiliation{3. Physikalisches Institut, University of Stuttgart, Germany, Research Center Scope and IQST}
\author{I. Jakobi}
\affiliation{3. Physikalisches Institut, University of Stuttgart, Germany, Research Center Scope and IQST}
\author{F. Dolde}
\affiliation{3. Physikalisches Institut, University of Stuttgart, Germany, Research Center Scope and IQST}
\author{C. Burk}
\affiliation{3. Physikalisches Institut, University of Stuttgart, Germany, Research Center Scope and IQST}
\author{R. Reuter}
\affiliation{3. Physikalisches Institut, University of Stuttgart, Germany, Research Center Scope and IQST}
\author{G. Waldherr}
\affiliation{3. Physikalisches Institut, University of Stuttgart, Germany, Research Center Scope and IQST}
\author{J. Honert}
\affiliation{3. Physikalisches Institut, University of Stuttgart, Germany, Research Center Scope and IQST}
\author{T. Wolf}
\affiliation{3. Physikalisches Institut, University of Stuttgart, Germany, Research Center Scope and IQST}
\author{A. Brunner}
\affiliation{3. Physikalisches Institut, University of Stuttgart, Germany, Research Center Scope and IQST}
\author{J.H. Shim}
\affiliation{Experimentelle Physik III, Universit\"at Dortmund, Germany}
\author{D. Suter}
\affiliation{Experimentelle Physik III, Universit\"at Dortmund, Germany}
\author{H. Sumiya}
\affiliation{Sumitomo Electric Industries, Ltd., Itami, Japan}
\author{J. Isoya}
\affiliation{Research Center for Knowledge Communities, University of Tsukuba, Tsukuba, 305-8550 Japan}
\author{J. Wrachtrup}
\affiliation{3. Physikalisches Institut, University of Stuttgart, Germany, Research Center Scope and IQST}

\begin{abstract}
Measuring local temperature with a spatial resolution on the order of a few nanometers has a wide range of applications from semiconductor industry over material to life sciences \cite{yue_nanoscale_2012}.
When combined with precision temperature measurement it promises to give excess to small temperature changes caused e.g. by chemical reactions or biochemical processes \cite{okabe_intracellular_2012}.
However, nanoscale temperature measurements and precision have excluded each other so far owing to the physical processes used for temperature measurement of limited stability of nanoscale probes \cite{vetrone_temperature_2010}.
Here we experimentally demonstrate a novel nanoscale temperature sensing technique based on single atomic defects in diamonds.
Sensor sizes range from millimeter down to a few tens of nanometers.
Utilizing the sensitivity of the optically accessible electron spin level structure to temperature changes \cite{acosta_temperature_2010} we achieve a temperature noise floor of $5\,$mK/$\sqrt{\mathrm{Hz}}$ for single defects in bulk sensors.
Using doped nanodiamonds as sensors yields temperature measurement with $130\,$mK/$\sqrt{\mathrm{Hz}}$ noise floor and accuracies down to $1\,$mK at length scales of a few ten nanometers.
The high sensitivity to temperature changes together with excellent spatial resolution combined with outstanding sensor stability allows for nanoscale precision temperature determination enough to measure chemical processes of few or single molecules by their reaction heat even in heterogeneous environments like cells. \\
\end{abstract}

\maketitle

\begin{figure*}
  \includegraphics[width=1.0\textwidth]{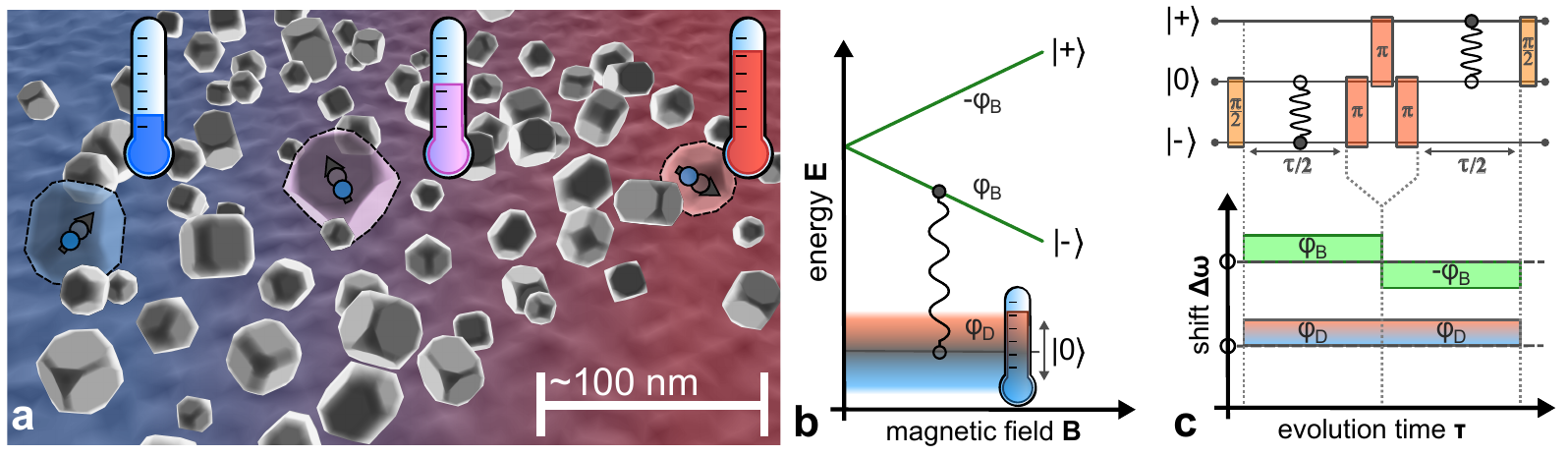}%
  \caption{\label{fig:1}
			\textbf{NV thermometer measurement scheme.}
			(a) Nanodiamonds containing single NV centers can serve as distributed probe temperature sensor.
			(b) Electron spin energy levels are mainly influenced by axial magnetic field $B_z$ and temperature $\Temp$.
			$B_z$ splits levels $\left| \pm \right\rangle$ and temperature shifts level $\left| 0 \right\rangle$ with respect to $\left| \pm \right\rangle$.
			During free evolution the spin states acquire phases $\varphi_D$, $\varphi_B$ and $-\varphi_B$.
			The initial superposition state $\left| 0 \right\rangle + \left| - \right\rangle$ of our decoupling sequence is depicted.
			(c) $D$-Ramsey sequence, upper part shows mw pulses for spin control and resulting spin states, lower part shows phase accumulation.
			}%
\end{figure*}

\begin{figure*}
  \includegraphics[width=1.0\textwidth]{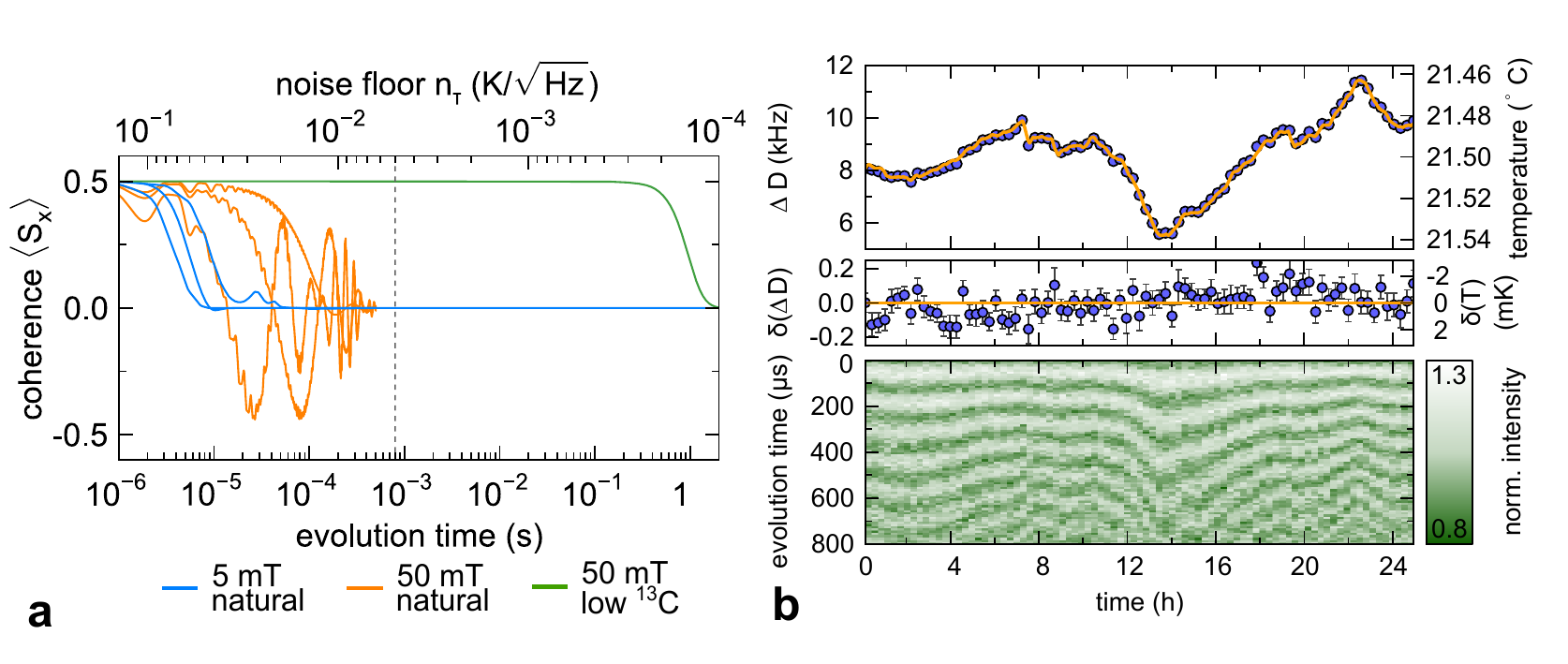}%
  \caption{\label{fig:2}
			\textbf{NV bulk diamond temperature sensor.}
			(a) Simulation of $D$-Ramsey coherence decays for $^{13}$C concentration $c=0.01$ and $B_z=5\,$mT (blue), for $c=0.01$ and $B_z=50\,$mT (orange) and for $c=10^{-5}$ and $B_z=50\,$mT
			(green).
			Only decoherence due to nuclear spin bath has been considered.
			(b) Bottom part shows consecutive $D$-Ramsey oscillations (vertical slices) of a single NV center.
			Horizontal and vertical axis are free evolution time $\tau$ total time $t$ respectively.
			The fluorescence response is color coded.
			Upper part shows the crystal field parameter $D=2870.685\,\mathrm{MHz}+\Delta D$ deduced from the $D$-Ramsey oscillations (circles).
			In addition the current temperature $\Temp$ of the diamond is plotted (yellow curve).
			We obtain a temperature dependence $c_{\Temp}=-78.6\pm 0.5\,\mathrm{kHz/K}$.
			Middle part shows difference between NV temperature results and additionally measured temperature ($\delta \left( \Delta D \right),\, \delta \Temp$).
			}%
\end{figure*}

\begin{figure*}
  \includegraphics[width=1.0\textwidth]{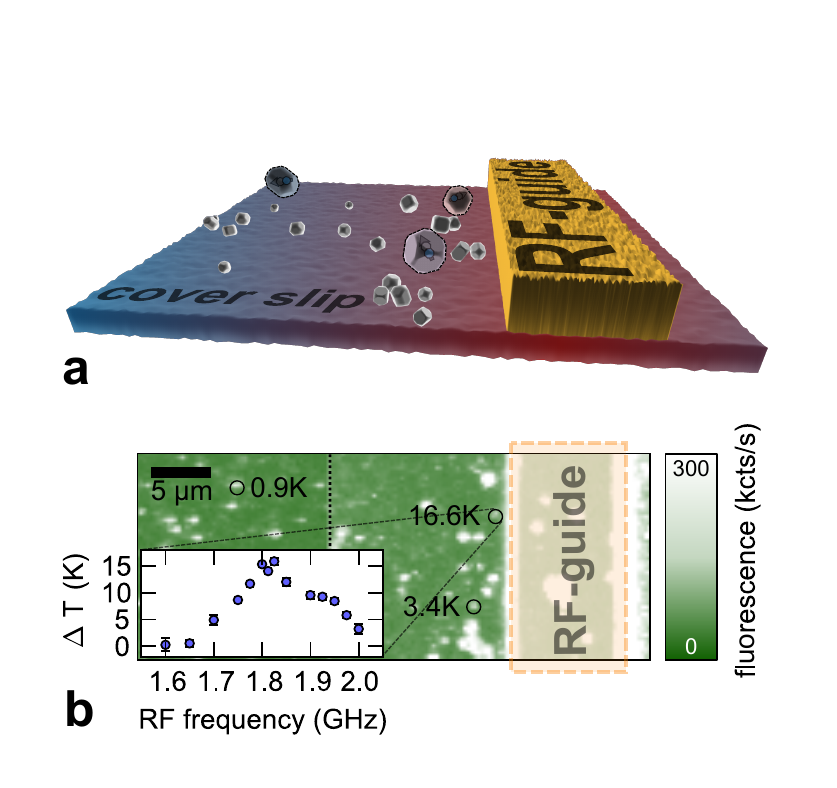}%
  \caption{\label{fig:3}
			\textbf{Nanodiamonds as temperature probes.}
			(a) Schematics of the experiment. Nanodiamonds scattered across a glass cover slide. A microstructured loop-gap resonator can be heated by rf radiation. 
			In the region of the current antinode the resonator dissipates energy due to ohmic loss, heating the underlying glass slide.
			Confocal imaging was performed from below the glass slide. 
			(b) Lateral confocal microscopy scan showing individual NV centers and part of the loop-gap resonator.
			The local temperature has been measured for three different distances from the resonator (black circles).
			(insert) Temperature shift of a single nanodiamond close to the wire in dependence of the rf frequency for resonator excitation.
			The resonance frequency of the resonator is revealed.}
\end{figure*}

Several kinds of nanoscale temperature sensing techniques have been developed in the recent past \cite{yue_nanoscale_2012}.
These are scanning thermal microscopes (SThM) \cite{majumdar_scanning_1999}, dispersed or scanned individual nanoprobes \cite{vetrone_temperature_2010,aigouy_scanning_2005}, direct methods like micro-Raman spectroscopy \cite{beechem_invited_2007} or near-field optical temperature measurements \cite{christofferson_microscale_2008}.
SThMs have temperature sensitive elements at a scanning tip (e.g. thermocouple), the nanoprobes have temperature dependent properties (e.g. fluorescence spectrum) which can be accessed without direct contact.
\\
\indent
In this study utilize a single quantum system in a solid state matrix as a temperature nanoprobe, namely the negatively charged nitrogen-vacancy (NV) center in diamond, which allows probe sizes down to $\sim 5\,$nm \cite{tisler_fluorescence_2009}.
High fidelity control of its ground state electronic and nuclear spins has been demonstrated for various quantum information test experiments \cite{jelezko_observation_2004,dutt_quantum_2007,neumann_multipartite_2008,shi_room-temperature_2010,neumann_single-shot_2010,dolde_room-temperature_2013} as well as for nanometer scale metrology purposes \cite{maze_nanoscale_2008,balasubramanian_nanoscale_2008,dolde_electric-field_2011,maletinsky_robust_2012} e.g. measuring small magnetic and electric fields.
Here we show that it also allows tracking temperature with high precision.
Temperature nanoprobes can be either dispersed in the specimen to be investigated or used in scanning probe geometry (see fig.~\ref{fig:1}a).
\\
\indent
The NV center is a molecular impurity in diamond comprising a substitutional nitrogen impurity and an adjacent carbon vacancy.
Optical excitation in a wavelength range from 460~nm to 580~nm yields intense fluorescence emission \cite{aslam_photo-induced_2013}.
Excitation also leads to a high degree of ground state electron spin polarization ($S=1$, the actual sensor level) into its $m_S=0\left(\left| 0 \right\rangle\right)$ sublevel \cite{waldherr_dark_2011}.
Furthermore the fluorescence decreases upon spin flips into the $m_S=\pm 1 \left(\left| \pm \right\rangle\right)$ sublevels allowing single spin optically detected magnetic resonance (ODMR) experiments at room temperature \cite{gruber_scanning_1997}.
Spins associated with the NV center possess long coherence times ($\sim 1\,$ms) \cite{neumann_multipartite_2008,mizuochi_coherence_2009} such that coherent control and metrology can be achieved with high fidelity.
A detectable temperature dependence of the spin resonance frequencies is observable down to temperatures of around 120~K, and very recently has been confirmed up to $700\,$K \cite{toyli_measurement_2012}.
\\
\indent
Optical excitation with 532~nm laser light and fluorescence collection of single NV centers is realized by a confocal microscope which images a bulk diamond sample or nanoscale diamonds.
Microwave (mw) radiation for spin manipulation are applied by appropriate wires close to the NV position.
The ODMR spectrum of the NV center spin usually comprises of two resonance lines corresponding to spin transitions between levels $\left| 0 \right\rangle \leftrightarrow \left| - \right\rangle$ and $\left| 0 \right\rangle \leftrightarrow \left| + \right\rangle$ \cite{gruber_scanning_1997} (see supporting information).
These spin levels depend on external parameters like magnetic ($B$) and electric fields, temperature $\Temp$ and strain (see fig.~\ref{fig:1}b).
By analyzing the spin Hamiltonian
\begin{eqnarray}
H &=& D(\Temp) S_z^2 + \gamma_{\mathrm{NV}} \left( B_z S_z + B_x S_x + B_y S_y\right) \nonumber \\
  &+& H_{\mathrm{hf}} + H_{\mathrm{nuc}}
\label{eq:Hamiltonian}
\end{eqnarray}
one can distinguish between individual contributions.
In Fig.~\ref{fig:1}b the first two largest terms of eq.~(\ref{eq:Hamiltonian}) are illustrated.
These are the crystal field which splits spin state $\left| 0 \right\rangle$ from states $\left| \pm \right\rangle$ and the Zeeman term due to axial magnetic field $B_z$ which splits states $\left| \pm \right\rangle$.
The latter terms commute.
The crystal field parameter $D$ depends on temperature $\Temp$, axial electric field and strain.
Under ambient conditions the temperature dependence is $c_{\Temp}=dD/d\Temp=-74.2\,\mathrm{kHz}/\mathrm{K}$ \cite{acosta_temperature_2010}.
By utilising a dedicated coherent control technique we reduced contributions from electric field and strain which are therefore neglected in eq.~(\ref{eq:Hamiltonian}).
The next smaller terms in the Hamiltonian is the transverse magnetic field ($B_{x},\,B_y$) which does not commute with the previous ones.
Transverse fields split levels $\left| \pm \right\rangle$ and additionally shift them with respect to $\left| 0 \right\rangle$. Therefore they can be confused with temperature changes.
Sufficiently strong axial magnetic fields suppress small transverse magnetic fields.
Eventually, each electron spin level is split by hyperfine interaction
\begin{equation}
H_{\mathrm{hf}} = \sum_{j=\mathrm{spins}}{\sum_{i=x,y,z}{S_z A_{zi}^j I_i^j}}
\label{eq:hf}
\end{equation}
to the nitrogen nuclear spin ($^{14}$N$ \rightarrow {I=1}$, $^{15}$N$ \rightarrow {I=1/2}$) \cite{rabeau_implantation_2006} and is inhomogeneously broadened by the $^{13}$C nuclear spin bath \cite{maze_electron_2008,mizuochi_coherence_2009}.
Other minor broadening effects are impurity electron spins or fluctuating electric or magnetic fields.
In diamonds with natural abundance of $^{13}$C the inhomogeneous ODMR linewidth is $1/T_2^*\sim 1\,$MHz.
This has severely limited temperature sensitivity in previous measurements where line shifts were used to measure temperature. \cite{acosta_temperature_2010,chen_temperature_2011,bouchard_detection_2011,toyli_measurement_2012}.
\\
\indent
Following the analysis of the spin Hamiltonian we adapt our experiments accordingly.
Non-commuting terms in eq.~(\ref{eq:Hamiltonian}), i.e. mainly transverse magnetic fields, are suppressed by a sufficiently large magnetic field aligned along the NV axis (i.e. $B=B_z$).
To suppress commuting parasitic effects, e.g. axial electric fields due to lattice defects, we use type IIa synthetic HPHT diamonds with low nitrogen content ($<\!14\,$ppb, see supporting information).
Additionally, we have developed a special decoupling sequence (see fig.~\ref{fig:1}c) which eliminates resonance line broadening due to magnetic interaction with e.g. spin impurities and at the same time is sensitive to changes in $D$ (i.e. temperature).
\\
\indent
As illustrated in fig.~\ref{fig:1}c we first create the superposition state $\left|0\right\rangle + \left|-\right\rangle$ which acquires a phase $\varphi_B-\varphi_D$ during time $\tau/2$, ($ e^{-\mathrm{i}\varphi_D}\left|0\right\rangle + e^{-\mathrm{i}\varphi_B}\left|-\right\rangle$), is then converted to $e^{-\mathrm{i}\varphi_D}\left|0\right\rangle - e^{-\mathrm{i}\varphi_B}\left|+\right\rangle$ and acquires an additional phase $-\varphi_B-\varphi_D$ during time $\tau/2$, ($e^{-\mathrm{i}2\varphi_D}\left|0\right\rangle - \left|+\right\rangle $).
The total phase $-2 \varphi_D = \Delta D \cdot \tau$ is proportional to shifts of $D$ only.
$\Delta D$ is a change of $D$ with respect to an initial resonance condition of spin levels and applied mw frequencies.
All intermediate phases $\varphi_B$ caused by quasi static fluctuations of magnetic field are canceled.
Hence, our decoupling sequence is of echo type for magnetic field fluctuations and of Ramsey type for changes in crystal field $D$ (i.e. $D$-Ramsey) with a coherence time $T_D$ of up to $\sim 1\,$ms corresponding to a homogeneous broadening of $1/T_D \sim 1\,$kHz.
Therefore we increase the phase accumulation time $\tau$ by orders of magnitude from $T_2^*$ to $T_D$ which in turn improves the frequency (temperature) uncertainty.
\\
\indent
The nuclear spin bath (see eq.~(\ref{eq:hf})) does not only lead to quasistatic fluctuating magnetic fields but additionally leads to entanglement between the bath spins and the electron spin \cite{childress_coherent_2006,maze_electron_2008}.
The latter effect also appears during echo sequences and effectively leads to faster decoherence.
The main requirement for this entanglement is that $A_{zx}$ and $A_{zy}$ terms of the hyperfine interaction are comparable with the nuclear Zeeman energy (see $H_{\mathrm{hf}}$ and $H_{\mathrm{nuc}}$ in eq.~(\ref{eq:hf}) and the supporting information).
We face this obstacle by reducing the $^{13}$C concentration for the HPHT bulk diamond samples and by increasing the magnetic field which reduces the number of bath spins for which the above mentioned requirement is fulfilled.
The effect of these two countermeasures can be estimated by numerically simulating the $D$-Ramsey decay due to the spin bath using the cluster expansion method \cite{maze_electron_2008} (see supporting information).
Figure~\ref{fig:2}b shows the expected $D$-Ramsey decay for a few magnetic field and $^{13}$C concentration settings.
\\
\indent
Our experiments are divided into two parts.
In the first part we test our novel $D$-Ramsey sequence for single NV centers in bulk diamond, to explore the best temperature sensitivity achievable. 
In the second part we apply NV centers in nanodiamonds as nanoscale sensors to measure temperature changes as a function of distance to a microstructure used as local heat source.
\\
\indent
In the first part (see fig.~\ref{fig:2}a) %
we consecutively measured full $D$-Ramsey oscillations for one day to track temperature and possible other effects on $D$.
Figure~\ref{fig:2}c shows the crystal field parameter $D$ calculated from the $D$-Ramsey oscillations' frequencies.
In addition the current temperature was measured by a thermistor close to the diamond (see supporting information) and for a value $c_{\Temp}=-78.6\pm 0.5\,\mathrm{kHz/K}$ both results coincide (see fig.~\ref{fig:2}c).
The average decay time of the $D$-Ramsey oscillations is $T_D = 829\pm 24\,\mu$s for measurement intervals of 17 minutes per point.
This corresponds to variations of the crystal field parameter $D$ of $\pm 0.2\,$kHz during one measurement interval.
Hence, temperature variations during one interval are at least within $\pm 2.5\,$mK.
Frequency uncertainties of fits to the $D$-Ramsey oscillations yield uncertainties for the average temperature during one interval of $1\,$mK (see error bars).
For the differences between NV based and external sensor based temperature data ($\delta(\Delta D)$ and $\delta \Temp$ in Scheme~\ref{fig:2}c) we obtain a standard deviation of $1\,$mK.
Our sensor therefore exhibits sufficient long term stability.
\\
\indent
Furthermore we have investigated the influence of laser and microwave heating during this measurement sequence by artificially increasing the heating power of laser and microwave.
Summarizing the microwave radiation has a negligible effect of much less than 1~mK while the laser heats the sample by less than 3~mK on average.
\\
\indent
Figure \ref{fig:2}c shows full $D$-Ramsey oscillations.
Highest temperature sensitivity, however, is only achieved for longest phase accumulation times $\tau$.
Taking into account only measurements with $\tau=800\,\mu$s we have achieved a temperature noise floor of $n_{\Temp} \approx 5\,\mathrm{mK}/\sqrt{\mathrm{Hz}}$ with a maximum measurement rate $r \approx 1\,$Hz.
For a given $r$ the uncertainty of each measurement point is $\sigma_{\Temp}=n_{\Temp} \sqrt{r}$.
Noise floor and the maximum measurement rate depend on $\tau$ approx. as $\sqrt{1/\tau}$ and $1/\tau$ respectively.
\\
\indent
In a second set of of our experiments utilizes NV centers in nanodiamonds.
We prepared a glass cover slide with a microstructured loop-gap resonator with a resonance frequency of $\approx 1.8\,$GHz and a linewidth of $\approx 200\,$MHz (see fig.~\ref{fig:3}a) which can be excited from a distance via radiofrequency (rf) radiation.
Afterwards nanodiamonds with a nominal size of 50~nm and with naturally occuring NV centers (see supporting information) were put in polyvinylalcohol and then spin-coated on top of this slide (see fig.~\ref{fig:3}a).
Single NV centers were addressed by a confocal microscope and their temperature response was investigated.
\\
\indent
Checks of the $T_2$ time and the $D$-Ramsey lifetime of the nanodiamonds yielded values in the range of 1 to $5\,\mu$s.
Next we heated the sample uniformly to check for the homogeneity of the temperature scaling $c_{\Temp}$ among the nanodiamonds.
For several tens of NV nanodiamonds we found slight variations corresponding to our measurement accuracy.
Next we applied our NV nanoscale thermometers to confirm heating by the micro structure.
To this end we measured temperature changes close to the microstucture wire using a single NV center in a nanodiamond (see fig.~\ref{fig:3}b).
When increasing the distance from the nanodiamond to the micro resonator by a few micrometers the temperature drops until ambient temperature is reached (see fig.~\ref{fig:3}c).
\\
\indent
The temperature noise floor reached in the nanodiamond experiment is $n_{\Temp} \approx 130\,\mathrm{mK}/\sqrt{\mathrm{Hz}}$, i.e. larger than for the bulk sample because of the shorter $T_D$ value.
This however allows for faster maximum measurement rates of $r\approx 100\,$Hz.
\\
\indent
Utilising our novel decoupling scheme nanoscale temperature sensing method results in a significant increase in temperature sensitivities using NV diamond.
We prove the long term accuracy of the sensor to within $1\,$mK.
In addition we demonstrate the applicability of NV nanodiamonds for temperature sensing with similar sensitivities recently reported bulk samples \cite{toyli_measurement_2012}.
Such nanodiamonds can be used as distributed temperature probes to reveal spatially resolved temperature profiles with spatial resolution only limited by their size.
\\
\indent
Temperature sensitivity can be further improved by a higher photon collection efficiency, e.g. by solid immersion lenses \cite{marseglia_nanofabricated_2011} or diamond pillar structures \cite{babinec_diamond_2010}.
In addition $T_D$ can be improved by analogues of higher order dynamical decoupling sequences like CPMG or UDD \cite{lange_universal_2010} allowing for micro Kelvin temperature sensitivity.
\\
\indent
Our results pave the way for implementing NV diamond temperature sensors for material and life sciences.
One might imagine the integration of NV centers into diamond AFM cantilever tips \cite{maletinsky_robust_2012}.
As distributed probes the selective attachment of nanodiamonds to specific parts of living cells or material and device parts seems feasible \cite{krueger_functionality_2012}.
While the temperature dependence of $D$ vanishes for temperatures below about $120\,$K other temperature dependent properties of the NV center, like the the zero-phonon line position of the fluorescence light, might become accessible at lower temperatures.
\\
\indent
For temperature measurements with higher time resolution pump probe measurements can be performed.
To this end the timing of heat generation and measurement sequence need to be properly synchronized.
As diamond has one of the highest heat conductivities fast response is guaranteed.
\\
\indent
We thank Fedor Jelezko, Ilja Gerhardt and Julia Michl for discussion.
We would like to acknowledge financial support by DFG and JST (SFB-TR\,21, FOR\,1482), EU (DIAMANT) and Baden-W\"urttemberg Stiftung gGmbH (``Methoden f\"ur die Lebenswissenschaften'').
A. Brunner acknowledges support by the Hans L. Merkle-Stiftung.

\bibliography{RefsZotero}

\end{document}